\documentclass[aps,pra,superscriptaddress,amsmath,amssymb,preprintnumbers,twocolumn]{revtex4}

\usepackage{amssymb}
\usepackage{graphicx}
\usepackage{graphics}
\usepackage{bm}
\usepackage{color}
\usepackage{cancel}
\usepackage{slashed}
\usepackage{subfigure}

\newcommand{\Ignore}[1]{}

\newcommand{\Ket}[1]{\left\vert #1\right\rangle}
\newcommand{\Bra}[1]{\left\langle #1\right\vert}

\newcommand{\BraKet}[2]{\left\langle#1\vert #2\right\rangle}

\newcommand{\ii}{\mathrm{i}}
\newcommand{\ee}{\mathrm{e}}

\begin{document}

\title{Synchronizing Quantum Harmonic Oscillators through Two-Level Systems}

\author{Benedetto Militello}
\address{Dipartimento di Fisica e Chimica, Universit\`a degli Studi di Palermo, Via Archirafi 36, I-90123 Palermo, Italy}
\address{I.N.F.N. Sezione di Catania}

\author{Hiromichi Nakazato}
\address{Department of Physics, Waseda University, Tokyo 169-8555, Japan}

\author{Anna Napoli}
\address{Dipartimento di Fisica e Chimica, Universit\`a degli Studi di Palermo, Via Archirafi 36, I-90123 Palermo, Italy}
\address{I.N.F.N. Sezione di Catania}

\begin{abstract}
Two oscillators coupled to a two-level system which in turn is coupled to an infinite number of oscillators (reservoir) are considered, bringing to light the occurrence of synchronization. A detailed analysis clarifies the physical mechanism that forces the system to oscillate at a single frequency with a predictable and tunable phase difference. Finally, the scheme is generalized to the case of $N$ oscillators and $M(<N)$ two-level systems.
\end{abstract}

\maketitle

\section{Introduction}\label{sec:introduction}

Synchronization processes in earlier definitions, imply the
alignment of the dynamics of two or more periodically evolving
physical systems~\cite{ref:Pikovski2003}. In classical mechanics
such processes are effectively described by the Kuramoto
model~\cite{ref_AcebronRMP2005}. Detailed studies, from basic
physical laws, of the occurrence of synchronization in classical
systems have been reported, especially considering pendula and
metronomes~\cite{ref_MaiantiAJP2009,ref_PantaleoneAJP2002}.
Synchronization phenomena turn out to be relevant in several
contexts, from neurosciences to medicine~\cite{ref_Strogatz,
ref_AngeliniPRE2004}, which has naturally brought to the extension
of the definition of synchronization to the dynamical alignment of
complex and even chaotic systems~\cite{ref_Arenas,ref:Padmanaban}.

Since the occurrence of synchronization has acquired attention in
the realm of nanotechnologies, attempts to extend the Kuramoto
model to the quantum realm have been
made~\cite{ref_deMendozaPRE2014}. Beyond the Kuramoto model,
studies of synchronization in quantum systems have been
developed~\cite{ref_ZambriniPRA2010,ref_ZambriniPRA2012,ref_ZambriniSciRep2013,ref_StrogatzPRL2004}.

Because of the lack of trajectories in quantum mechanics, giving a
proper definition of quantum synchronization is not as easy as in
classical mechanics. Mutual information and specific correlations
have been used, not only as a signature of dynamical alignment of
quantum systems, but also to evaluate the degree of
synchronization~\cite{ref_FazioPRL2013,ref_FazioPRA2015,ref_ArmourPRA2015}.

In the last years, the paradigm of quantum synchronization has
been extended from the first natural archetypical quantum system,
i.e. the harmonic oscillator, to the other fundamental class of
quantum systems, i.e. two-level systems (TLS). A few uncoupled
spins interacting with a common
environment~\cite{ref_Plastina2013} as well as ensembles of
dipoles~\cite{ref_Zhu2015} have been considered. More recently,
collective behavior of many spins has been studied in order to
establish a connection between synchronization processes and
superradiance or subradiance~\cite{ref_Bellomo2017}. A further
extension present in the literature is the dynamical alignment of
optomechanical
systems~\cite{ref_Li2016,ref_Li2017,ref_Vitali2017}.

Synchronization in hybrid systems, involving oscillators and TLSs
has been considered in the context of trapped ions in
Ref~\cite{ref_ArmourPRA2015}, where two oscillators (the centers
of mass of the ions) are considered as coupled to the
corresponding electronic degrees of freedom (modelled as two-level
systems) which are coupled to the electromagnetic field. Because
of the indirect interaction, the two oscillators eventually
synchronize. It is worth mentioning that the expression \lq
synchronization in hybrid systems\rq\, has not to be confused with
the \lq hybrid synchronization\rq\, present in the literature
which refers either to systems partially synchronized or to
different systems which align, in spite of their different
nature~\cite{ref:Padmanaban,ref_Qiu2015}. Moreover, the
possibility to drive dynamical alignment of oscillators (a driven
resonator) and TLSs (superconducting devices) has been predicted
and demonstrated~\cite{ref:Bistability2008,
ref:ArtificialAtom2007,ref:Bistability2017}. In particular, it has
been shown that the dynamical alignment bring to a significant
change of the qubit radiation spectrum~\cite{ref:Bistability2008,
ref:ArtificialAtom2007}

In this paper we will analyze a very simple hybrid model
consisting of two oscillators coupled to a single two-level
system, which in turn is coupled to an infinite number of
oscillators (reservoir), in order to study in depth the mechanism
of synchronization. We will show that the specific structure of
the couplings between the oscillators and the TLS determines the
structure of a preserved mode which in the long-time dynamics is
responsible for a collective motion characterized by a single
frequency. More detailed predictions about the phase difference
between the oscillations can be made, on the basis of knowledge of
the initial condition and of specific phase relations between the
coupling constants of the oscillators with the TLS. Also the
amplitudes of the position oscillations are easily predictable.
The essential features of this model could be extended to its
natural generalization, i.e. to the case of more than two
oscillators and many two-level systems.

It is worth emphasizing that the relevance of our model is
twofold. On the one hand, it shows that two \lq big\rq\, systems
(two oscillators, whose Hilbert spaces are infinite dimensional)
can be synchronized by the action of a single \lq small\rq\, one
(a TLS), also providing an example of how two similar systems can
be synchronized by a third part which has a completely different
nature. This very point, a small (finite-level) system, which does
not synchronize with the rest but which, at the same time, is
responsible for making all the rest (a bigger, infinite-level
system) synchronize, is a remarkable result. Of course, the key
ingredient is that the TLS is interacting with the environment. On
the other hand, the simplicity of the model (quantized oscillators
and TLSs with dipole-like interactions) makes it relevant in many
physical scenarios, from bimodal cavities to trapped ions to
superconducting devices interacting with quantized fields.
Moreover, such simplicity allows us to understand in a very clear
way the mechanism of synchronization, as well as to predict the
final common frequency of the oscillations, their amplitudes and
possible phase differences.

The paper is structured as follows. In Sec.\ref{sec:model} we
introduce the model and single out, qualitatively, some important
dynamical features, while in Sec.\ref{sec:evolutions} we show that
in the long-time dynamics, for every initial state of the system,
the two oscillators evolve with the same frequency. In
Sec.\ref{sec:generalization} we generalize the model to the case
of $N$ oscillators and $M(<N)$ TLSs. Finally, in
Sec.\ref{sec:conclusions} we provide conclusive remarks.

\section{The Model}\label{sec:model}

Consider a system governed by the following Hamiltonian:
\begin{subequations}
\begin{eqnarray}\label{eq:Haniltonian}
  \nonumber
  H_\mathrm{S} &=& \sum_{k=1}^2 \omega_k a^\dag_k a_k + \frac{\omega_0}{2}\sigma_z + \sum_{k=1}^2 g_k (\ee^{\ii\theta_k} a_k + \ee^{-\ii\theta_k} a_k^\dag ) \sigma_x \,, \\
\end{eqnarray}
and assume that the two-level system (TLS) is coupled to an
environment, here modelled as an infinite set of harmonic
oscillators, each one linearly coupled to the two-level system:
\begin{eqnarray}\label{eq:Haniltonian}
  H_\mathrm{B} &=& \sum_i \nu_i b_i^\dag b_i  \,, \\
  H_\mathrm{I} &=& \sum_i (\beta_i b_i + H.c. ) \sigma_x \,.
\end{eqnarray}
\end{subequations}
We will see in the following that this interaction with the
environment is crucial, since, in addition with the interaction
between the TLS and the oscillators, is responsible for the
appearance of dissipating modes and stable modes.

\subsection{\lq Single Leaking Mode Picture\rq\, and Preserved Mode}

Let us make a change of picture through the following unitary operators: a phase changing unitary operator,
 \begin{subequations}
\begin{eqnarray}\label{eq:UnitaryOp}
  U_p = \exp\left( \ii \theta_1 a_1^\dag a_1 + \ii \theta_2  a_2^\dag a_2 \right)\,,
\end{eqnarray}
which transforms $a_k$ into $\ee^{-\ii\theta_k} a_k$, and a rotation unitary operator,
\begin{eqnarray}\label{eq:UnitaryOp}
  U_r = \exp\left[ \gamma(a_1 a_2^\dag - a_1^\dag a_2) \right]\,,
\end{eqnarray}
which realizes the following transformation:
\begin{eqnarray}
   a_1 &\rightarrow& U_r a_1 U_r^\dag = \cos\gamma\,a_1 + \sin\gamma\, a_2  \,, \\
   a_2 &\rightarrow& U_r a_2 U_r^\dag = -\sin\gamma\, a_1 + \cos\gamma\, a_2  \,, \\
   \sigma_\alpha &\rightarrow& U_r \sigma_\alpha U_r^\dag = \sigma_\alpha \,, \qquad \alpha=x,y,z,\pm\,.
\end{eqnarray}
\end{subequations}
The parameter $\gamma$ determines how the two modes are
mixed, in this new picture (the Pauli operators are left unchanged
for any $\gamma$).

The system Hamiltonian transformed by $U_r U_p$ (but expressed in terms of the original operators, $a_k$'s and $\sigma_\alpha$'s) is given by:
\begin{subequations}
\begin{eqnarray}\label{eq:TransHam}
  \nonumber
  \tilde{H}_\mathrm{S} &=& \sum_{k=1}^2 \tilde{\omega}_k a^\dag_k a_k + \frac{\omega_0}{2}\sigma_z
     + \sum_{k=1}^2 \tilde{g}_k ( a_k + a_k^\dag) \sigma_x   \\
     &+& \left(\xi_{12} \, a_1^\dag a_2 + H.c. \, \right)  \,,
\end{eqnarray}
with
\begin{eqnarray}
  \tilde{\omega}_1 &=&  \omega_1 \cos^2\gamma + \omega_2\sin^2\gamma  \,, \\
  \tilde{\omega}_2 &=&  \omega_1 \sin^2\gamma + \omega_2\cos^2\gamma   \,, \\
  \tilde{g}_1 &=&   g_1\cos\gamma - g_2\sin\gamma \,, \\
  \tilde{g}_2 &=&   g_1\sin\gamma + g_2\cos\gamma \,, \\
  \xi_{12} &=& ( \omega_1-\omega_2 ) \sin\gamma \, \cos\gamma \,.
\end{eqnarray}
\end{subequations}
In this new picture, both the free energies of the
oscillators and the coupling strengths of the oscillators with the
TLS are modified. Moreover, because of the mixed algebra of the
bosonic ($a_k$'s) and fermionic ($\sigma_\alpha$'s) operators, it
is not possible to perfectly decouple the two modes, which implies
that a tunnelling term between the two oscillators appears, the
relevant strength being given by the parameter $\xi_{12}$.

By a suitable choice of the parameter $\gamma$,
\begin{eqnarray}
  \tan\gamma = \frac{g_1}{g_2}\,,
\end{eqnarray}
we obtain $\tilde{g}_1 = 0$ (and, by the way,
$\tilde{g}_2=\sqrt{g_1^2+g_2^2}\,$), which implies that in this
picture only one mode is directly coupled to the leaking TLS. (For
this reason we will call this picture the \lq single leaking mode picture\rq\,, or SLMP.) However, because of the coupling between the
two modes ($\xi_{12}\not=0$), the other mode has an indirect
coupling to the leaking TLS.

By imposing $\xi_{12} \ll \tilde{g}_2$ and $\xi_{12}$ much smaller than the decay rate of the TLS, we obtain a situation where the mode $1$ is essentially decoupled from the mode $2$ and the TLS, so that, in a certain time scale, the mode $2$ decays toward the ground state, and only the mode $1$ keeps some energy. Therefore, after a while, the whole system will oscillate at the frequency $\tilde{\omega}_1$.

The first condition ($\xi_{12} \ll \tilde{g}_2$) may be recast in
the following form:
\begin{eqnarray}\label{eq:suffcond1}
  |\omega_1-\omega_2| \ll \frac{(g_1^2+g_2^2)^{3/2}}{|g_1 g_2|} \,,
\end{eqnarray}
while the second condition is:
\begin{eqnarray}\label{eq:suffcond2}
  |( \omega_1-\omega_2 ) \sin\gamma \, \cos\gamma | \ll \Gamma \,,
\end{eqnarray}
where $\Gamma$ is the TLS decay rate.

The parameter $\xi_{12}$ can be put exactly equal to zero only in the following trivial cases: $\omega_1=\omega_2$ (bare modes already at the same frequency), $g_1=0$ ($\sin\gamma=0$, which means that one mode is not coupled to the dissipating TLS), $g_2=0$ ($\cos\gamma=0$, which means that the second mode is not coupled to the dissipating TLS).

The preserved mode is mode $1$, in this picture. When we come back to the original (Schr\"odinger) picture, it corresponds to:
\begin{eqnarray}
  \ee^{\ii\theta_1} \cos\gamma\,a_1 - \ee^{\ii\theta_2}\sin\gamma\, a_2 \propto \ee^{\ii\theta_1} g_2 a_1 - \ee^{\ii\theta_2} g_1 a_2 \,,
\end{eqnarray}
while the dissipating mode is
\begin{eqnarray}
  \ee^{\ii\theta_1}\sin\gamma\,a_1 + \ee^{\ii\theta_2}\cos\gamma\, a_2 \propto \ee^{\ii\theta_1} g_1 a_1 + \ee^{\ii\theta_2} g_2 a_2 \,,
\end{eqnarray}
which could be easily predicted. Indeed, the mode proportional to $\sum_k g_k \ee^{\ii\theta_k} a_k$ is the one involved in the interaction with the leaking TLS.

\subsection{Dissipative Dynamics}

As already pointed out, the system undergoes an evolution which is
essentially unitary for the mode $1$ and dissipative for mode $2$
and the TLS. In the SLMP, we can write:
\begin{eqnarray}\label{eq:MEMicro}
  \dot{\tilde{\rho}} = -\ii[\tilde{H}_\mathrm{S},
  \tilde{\rho}]+ {\cal D}_{2\sigma}\tilde{\rho}\,,
\end{eqnarray}
where ${\cal D}_{2\sigma}$ is the dissipator associated to the
subsystem made of the mode $2$ and the TLS.

In principle, the dissipator can be derived through standard
methods~\cite{ref:BreuerPetruccione,ref:Gardiner} and the dynamics
evaluated. Since the (sub)system is made of two parts interacting,
the interaction between the mode and the TLS should be considered
from the beginning~\cite{ref:PhenoVsMicro-1,ref:PhenoVsMicro-2},
and the dissipator is supposed to connect eigenstates of the
Hamiltonian (including the interaction). However, because of the
presence of the counter-rotating terms, it is not easy to deal
with such a problem beyond the perturbative
approach~\cite{ref:CRT1,ref:CRT2}. Anyway, since here we do not
want to develop a completely quantitative analysis, for our
purpose, it will be enough to consider that the eigenstates of the
Hamiltonian $\tilde{\omega}_2 a^\dag_2 a_2 + \omega_0/2\sigma_z +
\tilde{g}_2 ( a_2 + a_2^\dag) \sigma_x$ differ from those of the
RWA counterpart, $\tilde{H}_\mathrm{RWA}\equiv\tilde{\omega}_2 a^\dag_2 a_2 +
\omega_0/2\sigma_z + \tilde{g}_2 ( a_2 \sigma_+ + a_2^\dag
\sigma_-)$, with corrections of the order $\eta=g/\omega$, with
$g=\sqrt{g_1^2+g_2^2}$ and $\omega=\sqrt{\omega_1^2+\omega_2^2}$.

Since it is well known that synchronization phenomena occur after
a long time, it is reasonable to consider the dissipative dynamics
in the Markovian limit. Moreover, we assume low (virtually zero)
temperature for the environment, which implies that eventually the
dissipator will drive the relevant subsystem toward its lowest
energy state. Therefore, though it can appear a rough
approximation, we can reasonably assume that in the SLMP every state of the mode $2$ and the TLS eventually relaxes
toward the ground state of $\tilde{H}_\mathrm{RWA}$:
\begin{equation}
  \tilde\rho_{2\sigma} \rightarrow (1-\eta)\Ket{0}_2\Bra{0} \otimes \Ket{-}\Bra{-} +
  O(\eta)\,,
\end{equation}
where $\sigma_z|-\rangle=-|-\rangle$. On the other hand, every coherence/traceless operator eventually vanishes:
\begin{equation}
  \Ket{\psi_{2\sigma}}\Bra{\psi_{2\sigma}'} \rightarrow O(\eta)\,,
  \qquad \BraKet{\psi_{2\sigma}}{\psi_{2\sigma}'}=0\,.
\end{equation}

In order to make our predictions reliable, either we calculate the
corrections due to the counter-rotating terms or we assume
$\eta\ll 1$. In the second case (which is our choice, in this
paper), this means assuming,
\begin{equation}\label{eq:suffcond3}
  \eta=\sqrt{\frac{g_1^2+g_2^2}{\omega_1^2+\omega_2^2}} \ll 1\,,
\end{equation}
which must be added to the two conditions previously considered.

It is worth noting that essentially the same predictions come out from a phenomenological model, which consists in deriving the (zero-temperature) master equation for the TLS neglecting the coupling with the oscillators, i.e.:
\begin{eqnarray}\label{eq:MEPheno}
  \dot{\tilde{\rho}}_{2\sigma} = -\ii[\tilde{H}_\mathrm{S},
  \tilde{\rho}_{2\sigma}]+ {\cal D}_{\sigma}\tilde{\rho}_{2\sigma}\,,
\end{eqnarray}
with
\begin{eqnarray}
  {\cal D}_{\sigma}\tilde{\rho_{2\sigma}} = \Gamma (\sigma_- \tilde{\rho}_{2\sigma} \sigma_+ - 1/2\{\sigma_+\sigma_-, \tilde{\rho}_{2\sigma}\} )\,.
\end{eqnarray}
Indeed, the RWA counterpart of this model,
\begin{eqnarray}
  \dot{\tilde{\rho}}_{2\sigma} = -\ii[\tilde{H}_\mathrm{RWA},
  \tilde{\rho}_{2\sigma}]+ {\cal D}_{\sigma}\tilde{\rho}_{2\sigma}\,,
\end{eqnarray}
has $\Ket{0}_2\Ket{-}$ as a stationary state and describes processes which make every state $\tilde{\rho}_{2\sigma}$ eventually reach the ground state $\Ket{0}_2\Bra{0}\otimes \Ket{-}\Bra{-}$. (In fact, every state $\Ket{n}_2\Ket{+}$ will relax toward $\Ket{n}_2\Ket{-}$, and every state $\Ket{n}_2\Ket{-}$, with $n>0$, will undergo coherent transitions toward $\Ket{n-1}_2\Ket{+}$, which will relax toward $\Ket{n-1}_2\Ket{-}$). When we include the counter rotating terms, we just add corrections of the order of $\eta$.

It is worth mentioning that when $\tilde{H}_\mathrm{RWA}$ is
considered in place of $\tilde{H}_\mathrm{S}$, hence suppressing
the counter rotating terms, relevant shifts (Bloch-Siegert's)
should be considered. However, such shifts do not have a
significant role in our analysis, because whether they are taken
into account or neglected, the ground state is always
$\Ket{0}_2\Ket{-}$ and the tendency of the system to relax toward
such ground state is always present.

We conclude this section by summarizing the conditions that have
to be satisfied in order to guarantee the occurrence of
synchronization. Such conditions are given by
Eqs.(\ref{eq:suffcond1}), (\ref{eq:suffcond2}) and
(\ref{eq:suffcond3}). Altogether they require, more or less, that
the natural frequency difference $|\omega_1-\omega_2|$ is much
smaller than the coupling constants $g_1$ and $g_2$ as well as
much smaller than the natural decay rate $\Gamma$; in turn, all
such quantities are supposed to be much smaller than the natural
frequencies $\omega_1$ and $\omega_2$. It is also important to
remind that such conditions are only sufficient, not necessary, so
that we can have synchronization even out of the parameter regions
defined by them.

\section{Evolutions}\label{sec:evolutions}

Starting from the analysis of the previous section, we will be in
a condition to forecast the approaching of the system toward a
dynamical stationary state which exhibits the features of a
synchronized state. In the next two subsections we will consider
two very special initial conditions, then an arbitrary initial
condition will be considered and the statement about the
occurrence of synchronization will be generalized.

\subsection{Single-mode Coherent State}\label{sec:singlecoherent}

Let us start by considering the case where the system is prepared in a coherent state of one of the two oscillators (for example the first one):
\begin{eqnarray}
  \Ket{\psi (0)} = \Ket{\alpha_1}_1 \Ket{0}_2 \Ket{g} = D_1(\alpha_1) \Ket{0}_1 \Ket{0}_2 \Ket{-}\,,
\end{eqnarray}
with $\sigma_z\Ket{\pm}=\pm\Ket{\pm}$ and
\begin{eqnarray}
  D_k(\alpha_k) = \exp(\alpha_k a_k^\dag - \alpha_k^* a_k) \,.
\end{eqnarray}

After applying the unitary operator $U_r U_p$, we get:
\begin{eqnarray}
   \nonumber
   |\tilde{\psi} (0)\rangle &=& U_r U_p D_1(\alpha_1) \Ket{0}_1 \Ket{0}_2 \Ket{-} \\
   %\nonumber
   %&=& U_r U_p D_1(\alpha_1) U_p^\dag U_r^\dag U_r U_p \Ket{0}_1 \Ket{0}_2 \Ket{-} \\
   %\nonumber
   %&=& U_r U_p  D_1(\alpha_1) U_p^\dag U_r^\dag \Ket{0}_1 \Ket{0}_2 \Ket{-} \\
   &=& D_1(\tilde{\alpha}_1) D_2(\tilde{\alpha}_2)  \Ket{0}_1 \Ket{0}_2 \Ket{-} \,,
\end{eqnarray}
with
\begin{eqnarray}\label{eq:alphatilde}
  \tilde{\alpha}_1 = \ee^{\ii\theta_1} \cos\gamma \, \alpha_1\,, \qquad \tilde{\alpha}_2 = \ee^{\ii\theta_1}\sin\gamma \, \alpha_1\,.
\end{eqnarray}

Because of the interaction with the leaking TLS, the mode $2$ in
this picture (SLMP) will lose energy
and the system will approach a stationary condition
described by:
\begin{subequations}
\begin{eqnarray}
  \tilde{\rho}(t) &\approx& (1-\eta) |\tilde{\psi} (t)\rangle\langle \tilde{\psi} (t)| + O(\eta)\,, \\
  \nonumber
  |\tilde{\psi} (t)\rangle &=& \exp(-\ii \tilde{\omega}_1 a_1^\dag a_1 t) D_1(\tilde{\alpha}_1) \Ket{0}_1\Ket{0}_2 \Ket{-} \\
    &=& D_1(\tilde{\alpha}_1 \ee^{\ii \tilde{\omega}_1 t}) \Ket{0}_1 \Ket{0}_2 \Ket{-} \,.
\end{eqnarray}
\end{subequations}

In the original (Schr\"odinger) picture the evolution is given by:
\begin{subequations}
\begin{eqnarray}
  \rho(t) &\approx& (1-\eta) |\psi (t)\rangle\langle \psi (t)| + O(\eta)\,, \\
  \nonumber
  \Ket{\psi (t)} &\approx&  D_1(\bar{\alpha}_1 \ee^{\ii \tilde{\omega}_1 t}) D_2(\bar{\alpha}_2 \ee^{\ii \tilde{\omega}_1 t}) \Ket{0}_1 \Ket{0}_2 \Ket{-}
  \,,\\
\end{eqnarray}
with
\begin{eqnarray}
  \label{eq:finalamplitudesa}
  \bar{\alpha}_1 &=& %\ee^{-\ii\theta_1}\cos\gamma \, \tilde{\alpha}_1 =
  \cos^2\gamma \, \alpha_1 \,,\\
  \bar{\alpha}_2 %&=& -\ee^{-\ii \theta_2} \sin\gamma \, \tilde{\alpha}_1  \\
  %\nonumber
  %&=& -\ee^{\ii (\theta_2-\theta_1)} \sin\gamma\cos\gamma \, \alpha_1 \\
  &=& \ee^{\ii [\pi-(\theta_2-\theta_1)]} \sin\gamma\cos\gamma \, \alpha_1 \,,
  \label{eq:finalamplitudesb}
\end{eqnarray}
\end{subequations}
which essentially (up to terms of the order of $\eta$) describes a
situation where both the oscillators oscillate at the same
frequency, but with a relative phase $\pi-(\theta_2-\theta_1)$
which depends on the phases present in the couplings of the
oscillators with the TLS.

\subsection{Single-mode Fock States}\label{sec:singlefock}

Consider the system prepared into a single
Fock state of one of the oscillators, say $\Ket{n}_1 \Ket{0}_2
\Ket{-}$. After the change of picture through $U_r U_p$ this state
is mapped to a linear combination of the following structure:
$\Ket{n}_1 \Ket{0}_2 \Ket{-} \rightarrow  \sum_{k=0}^n c_k
\Ket{n-k}_1 \Ket{k}_2 \Ket{-}$, corresponding to the density operator,
\begin{eqnarray}
  \nonumber
  \tilde{\rho}(0) = \sum_{kj} c_k c_j^* \Ket{n-k}_1\Bra{n-j} \otimes \Ket{k}_2\Bra{j} \otimes  \Ket{-}\Bra{-}\,. \\
\end{eqnarray}

Because of the interaction of the second mode with the environment, all the off-diagonal terms ($k\not=j$) will disappear and the diagonal terms will decay toward the ground state of the second mode, so that the system will approach a stationary state which is diagonal in the Fock basis:
\begin{widetext}
\begin{eqnarray}
  \tilde{\rho}(t) &\approx& (1-\eta)
  \times \sum_k |c_k|^2 \Ket{n-k}_1\Bra{n-k} \otimes  \Ket{0}_2\Bra{0} \otimes  \Ket{-}\Bra{-} + O(\eta)\,.
\end{eqnarray}
\end{widetext}

Coming back to the original (Schr\"odinger) picture, no time dependence will be introduced, the system will not exhibit any evolution and therefore no synchronization will be visible.

\subsection{Arbitrary Initial State (Coherent State basis)}\label{sec:arbitrary}

The results of the previous sections, and in particular those of
sec.\ref{sec:singlecoherent}, can be generalized and proven to be
valid essentially for every initial state of the system. In fact,
we will consider an arbitrary initial condition, which can always
be expressed as a superposition of coherent states, and prove that
the whole system will eventually oscillate with a single
frequency. The generic initial state of the two oscillators (and
of the TLS in its ground state) can be expanded in terms of their
coherent states:
%\begin{subequations}
\begin{eqnarray}
\nonumber
|\psi (0)\rangle&=&\int{d^2\alpha_1d^2\alpha_2\over\pi^2}A(\alpha_1,\alpha_2)|\alpha_1\rangle|\alpha_2\rangle |-\rangle\\
\nonumber
&=&\int{d^2\alpha_1d^2\alpha_2\over\pi^2}A(\alpha_1,\alpha_2)D_1(\alpha_1)D_2(\alpha_2)|0_{12}\rangle\,,\\
\end{eqnarray}
where $\Ket{0_{12}} = \Ket{0}_1 \Ket{0}_2 \Ket{-}$.

%and
%\begin{eqnarray}
%D_k(\alpha) = e^{\alpha a_k^\dagger-\alpha^* a_k},\quad k=1,2.
%\end{eqnarray}
%%so that $a_k|\alpha\rangle=a_kD_k(\alpha)|0\rangle=\alpha|\alpha\rangle$.
%\end{subequations}

\begin{widetext}

Let us go to the SLMP through $U_r U_p$:
\begin{subequations}
\begin{eqnarray}
\nonumber
|\tilde{\psi} (0)\rangle&=& U_r U_p |\psi (0)\rangle =\int{d^2\alpha_1d^2\alpha_2\over\pi^2}A(\alpha_1,\alpha_2) U_r U_p D_1(\alpha_1)D_2(\alpha_2) U_p^\dag U_r^\dag |0_{12}\rangle\,,\\
%\nonumber
&=& \int{d^2\alpha_1d^2\alpha_2\over\pi^2}A(\alpha_1,\alpha_2)
D_1(\tilde{\alpha}_1)D_2(\tilde{\alpha}_2) |0_{12}\rangle\,,\\
%\end{eqnarray}
%with
%\begin{eqnarray}
\tilde{\alpha}_1 &=& \alpha_1 \ee^{\ii\theta_1} \cos\gamma -
\alpha_2\ee^{\ii\theta_2}\sin\gamma \,, \\
\tilde{\alpha}_2 &=& \alpha_1 \ee^{\ii\theta_1} \sin\gamma +
\alpha_2\ee^{\ii\theta_2}\cos\gamma  \,.
\end{eqnarray}
\end{subequations}

We analyze the evolution of single operator
$\Ket{\tilde{\alpha}_2}\Bra{\tilde{\alpha}_2'}$ in the Fock basis, where it can be
written as:
\begin{equation}
  \Ket{\tilde{\alpha}_2}_2\Bra{\tilde{\alpha}_2'} = \ee^{-(|\tilde{\alpha}_2|^2+|\tilde{\alpha}_2'|^2)/2} \sum_{k,j} \frac{\tilde{\alpha}_2^k [(\tilde{\alpha}_2')^*]^j}{\sqrt{k!
  j!}}\Ket{k}_2\Bra{j}\,.
\end{equation}

Under the action of the dissipator, all the off diagonal terms
%($k\not=j$, which are not proper coherences, since they refer to two different states, $\Ket{\alpha_2}_2$ and $\Ket{\alpha_2'}_2$)
will disappear, while the diagonal terms will gradually lose population in advantage of the ground state. Therefore we will have
\begin{eqnarray}
  \nonumber
  \Ket{\tilde{\alpha}_2}_2\Bra{\tilde{\alpha}_2'} &\rightarrow& (1-\eta) \, \ee^{-(|\tilde{\alpha}_2|^2+|\tilde{\alpha}_2'|^2)/2} \sum_{k}
  \frac{(\tilde{\alpha}_2\tilde{\alpha}_2'^*)^k}{k!}\Ket{0}_2\Bra{0} + O(\eta)\,\\
  &=& (1-\eta) \, \ee^{\tilde{\alpha}_2\tilde{\alpha}_2'^* - (|\tilde{\alpha}_2|^2+|\tilde{\alpha}_2'|^2)/2} \Ket{0}_2\Bra{0} + O(\eta)\,.
\end{eqnarray}

After introducing
\begin{equation}
  P(\tilde{\alpha}_2, \tilde{\alpha}_2') = \ee^{\tilde{\alpha}_2\tilde{\alpha}_2'^* - (|\tilde{\alpha}_2|^2+|\tilde{\alpha}_2'|^2)/2} \,,
\end{equation}
and neglecting the terms of the order of $\eta$ for the sake of
simplicity, we can write the state the system will approach as
follows:
\begin{eqnarray}
%\nonumber
\rho(t) &\approx& \int{d^2\alpha_1 d^2\alpha_2 d^2\alpha_1'
d^2\alpha_2'\over\pi^4}A(\alpha_1,\alpha_2)
A^*(\alpha_1',\alpha_2') P(\tilde{\alpha}_2, \tilde{\alpha}_2')
D_1(\tilde{\alpha}_1 \ee^{\ii\tilde{\omega}_1t})
|0_{12}\rangle\langle 0_{12}| D_1^\dag(\tilde{\alpha}_1'
\ee^{\ii\tilde{\omega}_1 t})\,,
\end{eqnarray}
which, in general, does not correspond to a pure state.

In the Schr\"odinger picture we get the following:
\begin{subequations}
\begin{eqnarray}
\nonumber
\rho(t) &\approx& \int{d^2\alpha_1 d^2\alpha_2 d^2\alpha_1' d^2\alpha_2'\over\pi^4}A(\alpha_1,\alpha_2) A^*(\alpha_1',\alpha_2') P(\tilde{\alpha}_2,\tilde{\alpha}_2') \\
%\nonumber%
&\times& %
D_1(x \ee^{\ii\tilde{\omega}_1 t})%
D_2(y\ee^{\ii\tilde{\omega}_1 t}) \,\, |0_{12}\rangle\,%
\langle 0_{12}| %
D_2^\dag(y'\ee^{\ii\tilde{\omega}_1 t}) %
D_1^\dag(x' \ee^{\ii\tilde{\omega}_1 t})\, \\ %
%\end{eqnarray}
%\begin{eqnarray}
x&=&\alpha_1\cos^2\gamma-\alpha_2\ee^{\ii(\theta_2-\theta_1)}\sin\gamma\cos\gamma\,,\\
y&=&\alpha_2\sin^2\gamma-\alpha_1\ee^{-\ii(\theta_2-\theta_1)}\sin\gamma\cos\gamma\,,\\
x'&=&\alpha_1'\cos^2\gamma-\alpha_2'\ee^{\ii(\theta_2-\theta_1)}\sin\gamma\cos\gamma\,,\\
y'&=&\alpha_2'\sin^2\gamma-\alpha_1'\ee^{-\ii(\theta_2-\theta_1)}\sin\gamma\cos\gamma\,.
\end{eqnarray}
\end{subequations}

On this basis we can evaluate the expectation values of the
$a_k$'s in the Schr\"odinger picture. First of all observe that in
general $\langle 0| D^\dag(\alpha \ee^{\ii\omega t}) D(\beta
\ee^{\ii\omega t}) |0\rangle = \langle 0| D^\dag(\alpha) D(\beta)
|0\rangle $, which is time-independent. Second, it is easy to demonstrate that the following quantities,
\begin{eqnarray}
\nonumber
&\int& \frac{d^2\alpha_1' d^2\alpha_2' d^2\alpha_2 d^2\alpha_2}{\pi^4} \, \alpha_k \, A^*(\alpha_1',\alpha_2')A(\alpha_1,\alpha_2) P(\tilde{\alpha}_2,\tilde{\alpha}_2') \\
\nonumber%
&\times& \langle0_{12}|D_1(x e^{\ii\tilde\omega_1t})
D_2(ye^{\ii\tilde\omega_1t}) D_2^\dag(y' e^{\ii\tilde\omega_1t})
D_1^\dag(x' e^{\ii\tilde\omega_1t}) |0_{12}\rangle\,\\
\nonumber%
&=&\int \frac{d^2\alpha_1' d^2\alpha_2' d^2\alpha_2
d^2\alpha_2}{\pi^4} \, \alpha_k \,
A^*(\alpha_1',\alpha_2')A(\alpha_1,\alpha_2)
P(\tilde{\alpha}_2,\tilde{\alpha}_2') \,
\langle0_{12}|D_2^\dag(y') D_1^\dag(x') D_1(x) D_2(y)
|0_{12}\rangle\, \\
&=&
\int \frac{d^2\alpha_1' d^2\alpha_2' d^2\alpha_2
d^2\alpha_2}{\pi^4} A(\alpha_1,\alpha_2)
A^*(\alpha_1',\alpha_2')\, \alpha_k \, \times \nonumber \\
%\nonumber
%&\times&
%\ee^{\tilde{\alpha}_2\tilde{\alpha}_2'^*-(|\tilde{\alpha}_2|^2+|\tilde{\alpha}_2'|^2)/2}
%\ee^{-|x-x'|^2/2} \ee^{(x^*x'-xx'^*)/2} \ee^{-|y-y'|^2/2}
%\ee^{(y^*y'-yy'^*)/2}\,, \\
&\times&
\ee^{\tilde{\alpha}_2\tilde{\alpha}_2'^*-(|\tilde{\alpha}_2|^2+|\tilde{\alpha}_2'|^2)/2}\,
\ee^{x x'^*-(|x|^2+|x'|^2)/2} \, \ee^{y y'^*-(|y|^2+|y'|^2)/2} \,,
\end{eqnarray}
with $k=1,2$, when definitions of $\tilde{\alpha}_k$, $\tilde{\alpha}_k'$, $x$, $x'$, $y$, $y'$ are considered, turn out to be equal to the mean values of the annihilation
operators in the initial state:
\begin{eqnarray}
\nonumber
\langle a_k \rangle_0 &\equiv&
\Bra{\psi(0)} a_k \Ket{\psi(0)} = \int \frac{d^2\alpha_1' d^2\alpha_2' d^2\alpha_2
d^2\alpha_2}{\pi^4} \,
A^*(\alpha_1',\alpha_2')A(\alpha_1,\alpha_2)  \, \alpha_k \times \\
&\times&
\ee^{-|\alpha_1-\alpha_1'|^2/2}
\ee^{(\alpha_1\alpha_1'^*-\alpha_1'\alpha_1^*)/2}
\ee^{-|\alpha_2-\alpha_2'|^2/2}
\ee^{(\alpha_2\alpha_2'^*-\alpha_2'\alpha_2^*)/2}\,.
\end{eqnarray}

Following a straightforward calculation, we then obtain the following expressions for the mean values of $a_k$'s:
\begin{subequations}
\begin{eqnarray}\label{eq:amplitude1}
\langle a_1\rangle (t)&=&\ee^{\ii\tilde\omega_1t}(\langle a_1 \rangle_0 \cos^2\gamma+\langle a_2 \rangle_0 \ee^{-\ii[\pi-(\theta_2-\theta_1)]}\cos\gamma\sin\gamma)\,,
\,
\end{eqnarray}
\begin{eqnarray}
\label{eq:amplitude2}
\langle a_2\rangle (t)&=&\ee^{\ii\tilde\omega_1t}(\langle a_1 \rangle_0 \ee^{\ii[\pi -(\theta_2-\theta_1)]}\cos\gamma\sin\gamma+\langle a_2 \rangle_0\sin^2\gamma).
\end{eqnarray}
\end{subequations}
Notice that the quantities in the parantheses are time-independent complex
numbers. This means that the two oscillators finally oscillate
with the same frequency $\tilde\omega_1$ with a definite phase
difference, which is determined by the ratio and the phases of the
coupling constants to the environment and the initial condition.
This conclusion is  valid for an arbitrary initial state, that is,
$\forall A(\alpha_1,\alpha_2)$. Of course, there are states, like
Fock states, for which $\langle a_1\rangle_0=\langle a_2\rangle_0=0$, so that the oscillations at
the same frequency $\tilde{\omega}_1$ are not present.
Nevertheless, this is not to be intended as a lack of
synchronization, but as a lack of visibility of the phenomenon.

It is worth commenting, at this point, that the study of
synchronization is often based on numerical treatments (see for
example
Refs.~\cite{ref_ArmourPRA2015,ref:Bistability2017,ref:Bistability2008,ref_Qiu2015}),
even though more analytical studies are also
present~\cite{ref_Li2016, ref:Tilley}. Moreover, qualitative
predictions based on the study of the normal modes and the
individualization of leaking and protected ones are present in the
literature(see for example Ref.~\cite{ref_ZambriniSciRep2013}). A
direct connection between the initial state of the system and the
properties of the synchronized motion (amplitudes and phases) is
reported here. Our semi-quantitative approach has given us the
possibility of bringing to light such connection in a quite simple
way, allowing us to foresee, not only the joint frequency, but
also phase differences and amplitudes of the motions of the
oscillators. As we will see in the next section, the agreement
between our theoretical analysis and the numerical calculations is
very good.

\end{widetext}

\section{Numerical results}

\begin{widetext}

\begin{figure}
%\subfigure[]{\includegraphics[width=0.35\textwidth, angle=0]{fig1a.eps}} \qquad\qquad\qquad%
%\subfigure[]{\includegraphics[width=0.35\textwidth, angle=0]{fig1b.eps}} \\%
%\subfigure[]{\includegraphics[width=0.35\textwidth, angle=0]{fig1c.eps}} \qquad\qquad\qquad%
%\subfigure[]{\includegraphics[width=0.35\textwidth, angle=0]{fig1d.eps}} %
\subfigure[]{\includegraphics[width=0.35\textwidth, angle=0]{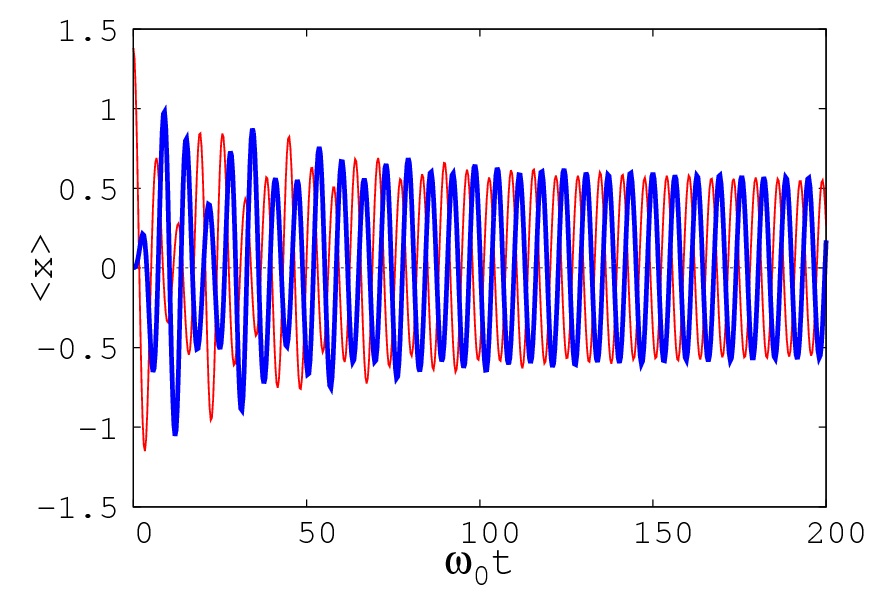}} \qquad\qquad\qquad%
\subfigure[]{\includegraphics[width=0.35\textwidth, angle=0]{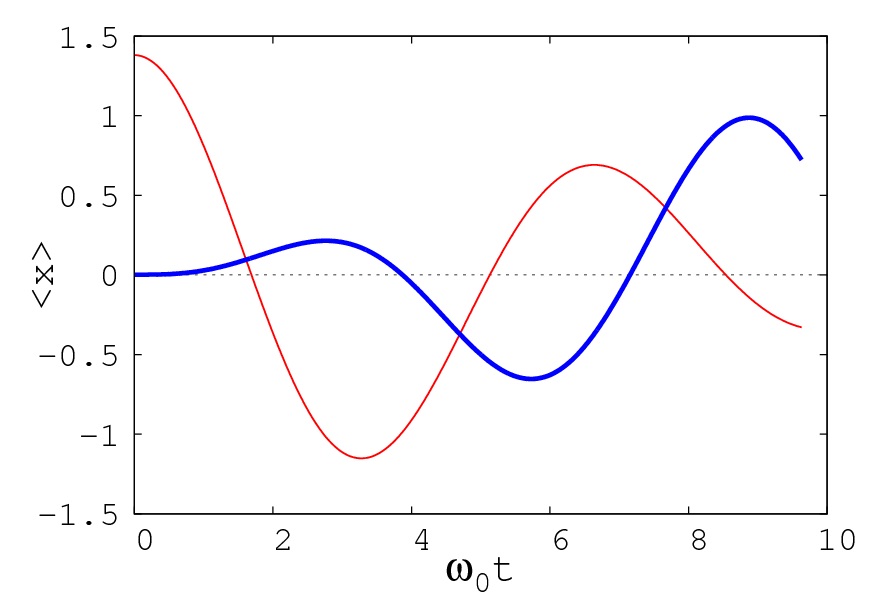}} \\%
\subfigure[]{\includegraphics[width=0.35\textwidth, angle=0]{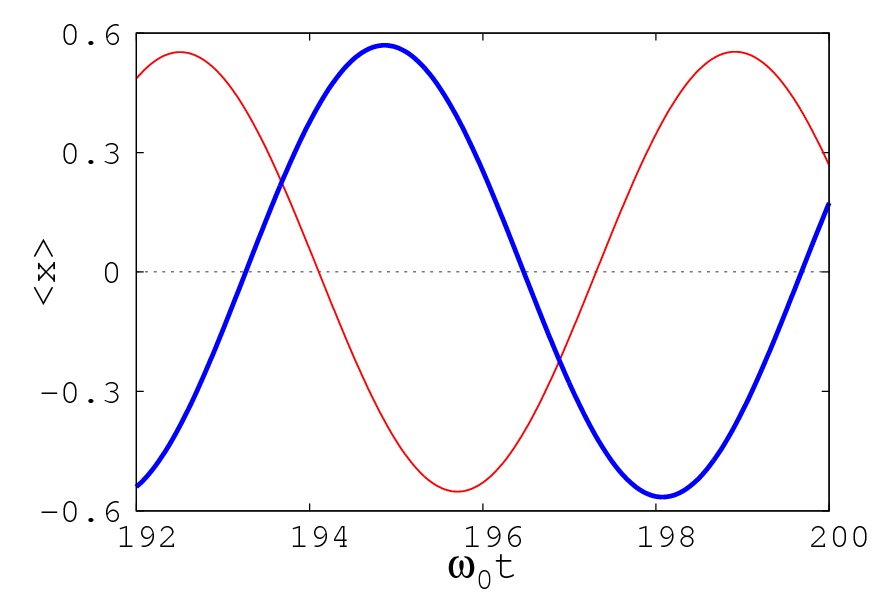}} \qquad\qquad\qquad%
\subfigure[]{\includegraphics[width=0.35\textwidth, angle=0]{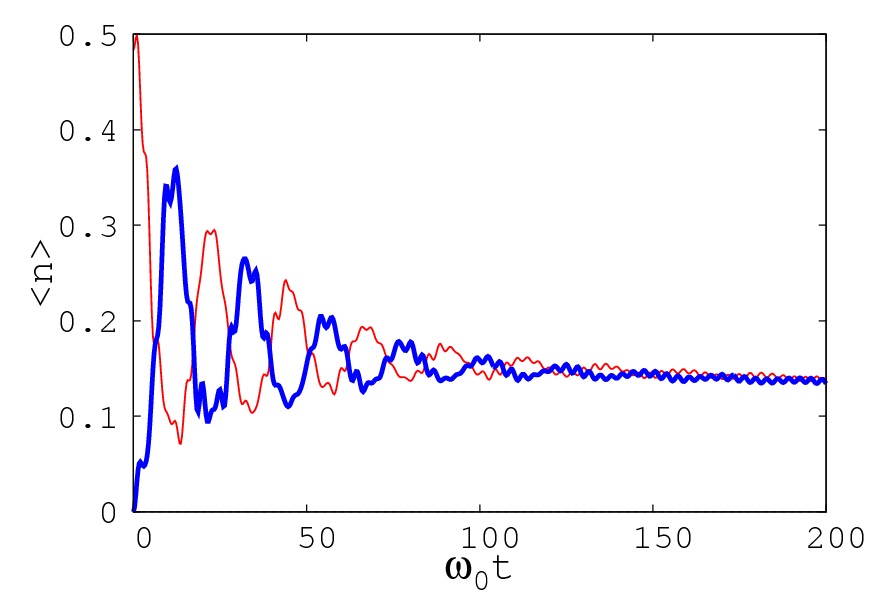}} %
\caption{(Color online) . Evolution of the system when the
oscillator 1 is prepared in a coherent state with $\alpha_1=0.7$
and the oscillator 2 is in the ground state; the TLS is in the
state $\Ket{-}$. The relevant parameters are:
$\omega_1/\omega_0=0.95$, $\omega_2/\omega_0=1.01$,
$g_1/\omega_0=0.2$, $g_2/\omega_0=0.21$, $\Gamma/\omega_0=0.1$,
$\theta_1=0$, $\theta_2=\pi/4$. We show the behavior of $\langle
x_1\rangle$  (thin red line) and $\langle x_2\rangle$  (bold blue
line), in a wide range (a), for short time (b) and in the final
part of the considered interval of time (c). It is clear that in
the beginning the expected values of the positions oscillate with
different frequencies, while in the final part of the evolution
they oscillate with the same frequency but with a phase difference
which is equal to $\pi-(\theta_2-\theta_1)=3\pi/4$, as expected
from the theory. In (d) it is also shown the average number of
excitations in each oscillator.} \label{fig:Evol_Pi4l}
\end{figure}
The results predicted by the previous theoretical analysis are
confirmed by our numerical simulations, which have been developed
considering a zero-temperature reservoir whose action is described
by the phenomenological model in \eqref{eq:MEPheno}. In
Fig.\ref{fig:Evol_Pi4l} we show the dynamics of the system when
one of the two oscillators is prepared in a coherent state (with a
small number of average excitations, in order to make possible
truncation of the Hilbert space at a reasonable point), the other
oscillator is in its vacuum and the TLS is in its ground state. We
consider the expectation values of the two positions $\langle
x_1\rangle$ and $\langle x_2\rangle$ in (a), (b) and (c), and the
average values of excitations $\langle n_1\rangle$ and $\langle
n_2\rangle$ in (d). It is well visible that after a certain time
the dynamics stabilizes, both in terms of amplitudes of the
oscillations of the positions (see (a)) and in terms of average
numbers of excitations (see (d)). A closer look at short-time and
long-time dynamics ((b) and (c), respectively) shows that the
positions initially oscillate with different frequencies, while
end up to oscillate at the same frequency and with a phase
difference which is $3\pi/4$, as expected from the theoretical
analysis, for the specific values of the parameters.

In Fig.\ref{fig:Evol_0_Pi} we find that the final situation can be
different (oscillations with the same frequency but phase
differences equal to $0$ or $\pi$), depending on the parameters.

Observe that in all such pictures the asymptotic oscillations have
almost the same amplitude. This is due to the fact that we have
$g_1/g_2\approx 1$, so that $\gamma\approx\pi/4$ (see
\eqref{eq:finalamplitudesa}-\eqref{eq:finalamplitudesb}). For the
same reason, the amplitudes of asymptotic oscillations of the two
modes are predicted to be essentially one half of the amplitude of
the initial oscillation of oscillator $1$ (see
\eqref{eq:amplitude1}-\eqref{eq:amplitude2}), which is well
visible in  Fig.\ref{fig:Evol_Pi4l}a.

\begin{figure}
%\subfigure[]{\includegraphics[width=0.35\textwidth, angle=0]{fig2a.eps}} \qquad\qquad\qquad%
%\subfigure[]{\includegraphics[width=0.35\textwidth, angle=0]{fig2b.eps}} %
\subfigure[]{\includegraphics[width=0.35\textwidth, angle=0]{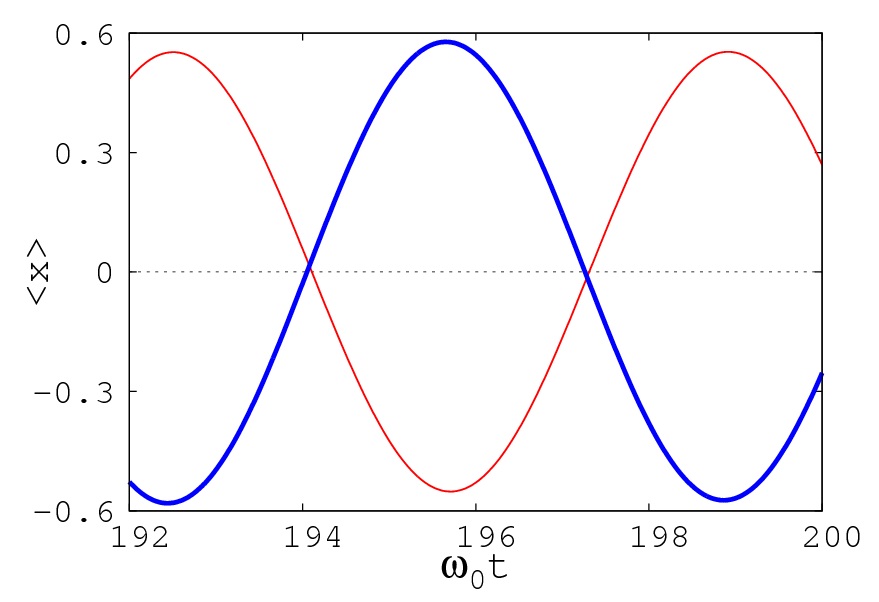}} \qquad\qquad\qquad%
\subfigure[]{\includegraphics[width=0.35\textwidth, angle=0]{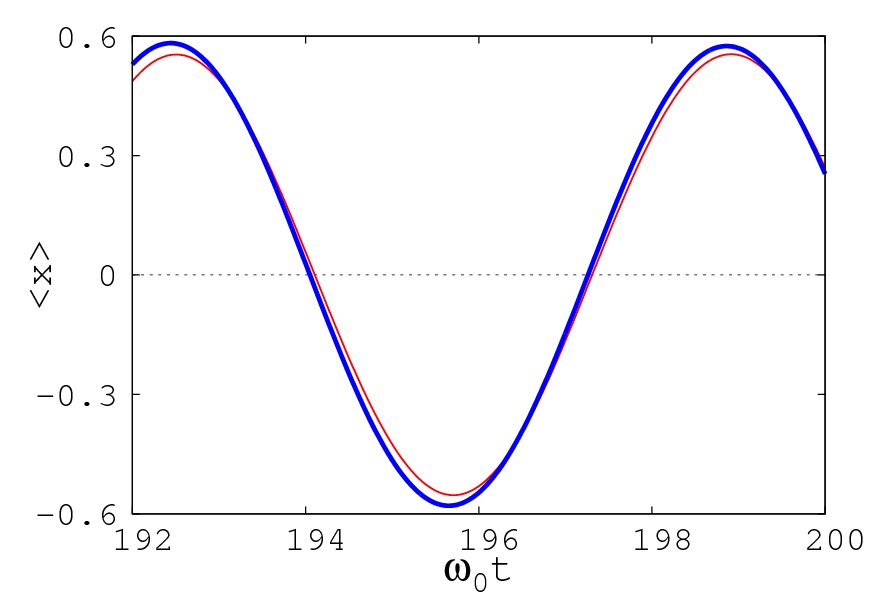}} %
\caption{(Color online) . Asymptotic single-frequency evolution of
the expectation values of the positions, with opposite phases (a)
and the same phase (b). The relevant parameters are:
$\alpha_1=0.7$, $\omega_1/\omega_0=0.95$,
$\omega_2/\omega_0=1.01$, $g_1/\omega_0=0.2$, $g_2/\omega_0=0.21$,
$\Gamma/\omega_0=0.1$. In (a) we have $\theta_1=0$, $\theta_2=0$,
while for (b) it is $\theta_1=0$, $\theta_2=\pi$. Thin red line
for the oscillator 1 and bold blue line for oscillator 2.}
\label{fig:Evol_0_Pi}
\end{figure}

\end{widetext}

\section{Generalizing the Model}\label{sec:generalization}

%\subsection{$N$ oscillators and $M$ TLS}

We can try to generalize the result obtained with a single TLS, by
using $M$ TLSs, each one \lq killing\rq\, a specific linear
combination of the bare modes. Consider the following Hamiltonian:
\begin{eqnarray}\label{eq:HaniltonianNxM}
  \nonumber
  H_\mathrm{S} &=& \sum_{k=1}^N \omega_k a^\dag_k a_k + \sum_{j=1}^M  \frac{\omega_0}{2} \sigma_{z}^j \\
                          &+& \sum_{j=1}^M \sum_{k=1}^N  g_{jk} (\ee^{\ii \theta_{jk}} a_k + \ee^{-\ii \theta_{jk}} a_k^\dag ) \sigma_{x}^j \,.
\end{eqnarray}

The following linear combinations of the bare modes are involved
in the interaction with the leaking TLSs:
\begin{eqnarray}\label{eq:HaniltonianNxM}
     c_j = G^{-1}\sum_{k=1}^N g_{jk} \ee^{\ii \theta_{jk}} a_k \,, \qquad j=1,..., M\,.
\end{eqnarray}
with $G=\sqrt{\sum_j |g_j|^2}$.

In general, they are not independent, but they can be generated as
linear combinations of $M$ independent modes. Let us call
$\bar{c}_i$, $i=1,..., N$, $N$ independent modes such that
$\bar{c}_{N-M+1},..., \bar{c}_N$, suitably combined, generate
$c_1,..., c_M$. The remaining $N-M$ modes, $\bar{c}_1,...,
\bar{c}_{N-M}$, are not coupled to the TLSs.

After a while, the modes related to  $\bar{c}_{N-M+1},...,
\bar{c}_N$ will waste all their energy and the dynamics will be
associated only to the modes $\bar{c}_1,..., \bar{c}_{N-M}$. If
the frequencies of the surviving modes are very close, the $N$
oscillators constituting the system will essentially evolve with a
single frequency.

\begin{figure}
%\subfigure[]{\includegraphics[width=0.35\textwidth, ,natwidth=610,natheight=642, angle=0]{fig3.pdf}}%
\includegraphics[width=0.35\textwidth, angle=0]{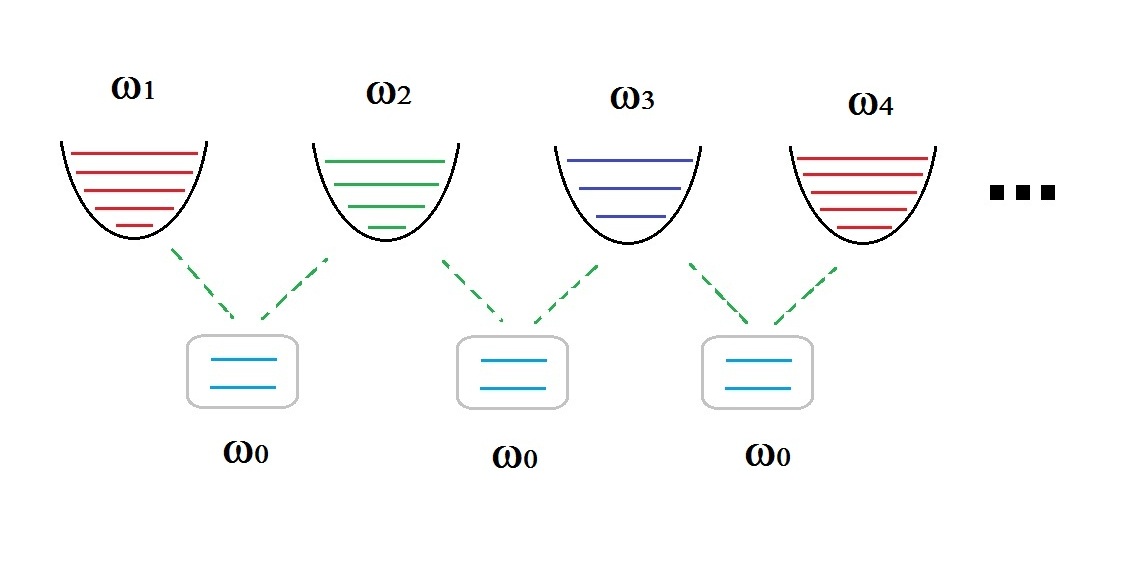}%
\caption{(Color online). A chain of $N$ oscillators coupled to a
chain of $N-1$ TLSs.} \label{fig:scheme}
\end{figure}

As a specific example, consider $N$ oscillators and
$M=N-1$ dissipating two-level systems, and assume that each couple
of adjacent oscillators is coupled to one of the TLS (see the
scheme in Fig.\ref{fig:scheme}). In particular, consider the
following Hamiltonian:
%\begin{eqnarray}\label{eq:Chain}
%  \nonumber
%  H_\mathrm{S} &=& \sum_{k=1}^N \omega_k a^\dag_k a_k + \sum_{j=1}^{N-1}  \frac{\omega_0}{2} \sigma_{z}^j \\
%                          &+& \sum_{j=1}^M g_{j} (\ee^{\ii \theta_{j}} a_j + \ee^{\ii \eta_{j+1}} a_{j+1} + h.c. ) \sigma_{x}^j \,.
%\end{eqnarray}
%
%Consider the special case,
\begin{eqnarray}\label{eq:Chain}
  \nonumber
  H_\mathrm{S} &=& \sum_{k=1}^N \omega_k a^\dag_k a_k + \sum_{j=1}^{N-1}  \frac{\omega_0}{2} \sigma_{z}^j \\
                          &+& \sum_{j=1}^M g_{j} (a_j - a_{j+1} + h.c. ) \sigma_{x}^j \,.
\end{eqnarray}

It is easy to see that it turns out that the mode
$a=(a_1+a_2+\cdots+a_N)/\sqrt{N}$ is protected, while the $N-1$ modes $(a_1-a_2)/\sqrt{2}$,
$(a_2-a_3)/\sqrt{2}$,..., $(a_{N-1}-a_N)/\sqrt{2}$ (which are not independent from each
other, but are all independent from $a$ and can be
rearranged to form a set of $N-1$ independent modes) are leaking.
Then, after a while, only mode $a$ will survive and we will
observe a collective motion with a single frequency and all the
oscillators with definite phase difference.

%\subsection{$N$ oscillators and a $M$-level system}

%It could be interesting to analyze the $M\rightarrow \infty$ limit, and check whether it gives results close to that of an harmonic oscillator.

\section{Conclusions}\label{sec:conclusions}

In this paper we have considered the occurrence of synchronization
of harmonic oscillators induced by two-level systems. Because of
the mixed algebra involving creation/annihilation operators on one
side and Pauli operators on the other side, it is not possible to
perfectly decouple the modes of the two oscillators from each
other and,  at the same time, one of them from the TLS.
Nevertheless, we have introduced a new picture, that we address
\lq single leaking mode picture\rq, where the two (transformed) modes are
almost decoupled from each other (provided a certain condition is
satisfied), and one of them is perfectly decoupled from the
leaking two-level system.  This circumstance allows us to forecast
with a good degree of reliability that one of the two modes will
lose energy, approaching the ground state, while the other will
evolve unitarily. This will produce, in the original
(Schr\"odinger) picture a collective oscillatory motion
characterized by a single frequency.

Our predictions based on the Hamiltonian model of the system are
corroborated by numerical results that show, in a clear way, not
only the incoming of a single frequency, but also specific phase
relations between the motions of the two oscillators predicted by
our theoretical analysis. Such phase relation is very easily
connected with the coupling constants ($g_k$'s) of the oscillators
with the TLS, when the system is prepared in a single-mode
coherent state. Even the amplitude of the oscillation of the
position is put in connection with both the relevant parameters
and the initial condition in a clear and simple way. Our
theoretical analysis provides also general expressions valid for
an arbitrary initial condition. A merit of our approach is that,
in spite of the semi-qualitative nature of our analysis (in the
sense that the dissipative dynamics is not solved or studied in
detail), it enables us to make quantitative predictions about the
joint frequency, individual amplitudes and phases of the
synchronized motions.

Finally, in sec.\ref{sec:generalization} we have provided a
generalization of the model involving $N$ oscillators and $M(<N)$
TLSs. By tuning the coupling constants between oscillators and
leaking TLSs it is possible to induce synchronization of all the
$N$ oscillators.

\section{Acknowledgements}

H. N. was supported by the Waseda University Grant for Special Research Projects (No. 2016B-173).

\end{document}